\documentclass[11pt]{article}
\usepackage{epsfig,amssymb,latexsym,color,amsmath,pifont}
\newcommand{\n}{\noindent} 
\newcommand{\e}{\varepsilon}
\newcommand{\x}{\mathbf{x}}

\begin{document}
\twocolumn[

\title{\textcolor{blue}{\textbf{Stability and chaos in coupled two-dimensional maps on Gene Regulatory Network of bacterium E.Coli}}}
\vspace{0.5cm}
\author{Zoran Levnaji\'c${}^{\; 1,2}$, Bosiljka Tadi\'c${}^{\; 2}$}
\date{}
\maketitle

\begin{center}
\begin{minipage}{6.46in}
${}^{1}$\textit{Department of Physics and Astronomy, University of Potsdam, Karl-Liebknecht-Street 24/25, D-14476 Potsdam-Golm, Germany} \\
${}^{2}$\textit{Department of Theoretical Physics, Jo\v zef Stefan Institute, Jamova 39, SI-1000 Ljubljana, Slovenia} \\ \\
The collective dynamics of coupled two-dimensional chaotic maps on complex networks is known to exhibit a rich variety of emergent properties which crucially depend on the underlying network topology. We investigate the collective motion of Chirikov standard maps interacting with time delay through directed links of Gene Regulatory Network of bacterium Escherichia Coli. Departures from strongly chaotic behavior of the isolated maps are studied in relation to different coupling forms and strengths. At smaller coupling intensities the network induces stable and coherent emergent dynamics. The unstable behavior appearing with increase of coupling strength remains confined within a connected sub-network. For the appropriate coupling, network exhibits statistically robust self-organized dynamics in a weakly chaotic regime.
\end{minipage}
\end{center} ] 


\n \textbf{The system of transcriptional regulations among genes in a cell is suitably modeled by Gene Regulatory Network, with nodes representing genes, and directed links modeling gene interactions. Dynamical state of an individual gene can be described by a pair of variables representing concentrations of protein and \textit{m}RNA with stationary time variations over the cell-cycle. The architecture of Gene Regulatory Networks contribute to cell's optimal functionality and dynamical stability, which are both necessary for its survival. An interesting open question regards the potentials of the structure of a given Gene Regulatory Network to stabilize the dynamics for a wider class of nonlinear dynamical systems, such as coupled chaotic maps. It has been recognized that chaotic maps on networks of different types display a variety of collective effects including self-organized stable dynamics and new types of attractors. In this paper we consider Gene Regulatory Network of bacterium Escherichia Coli and examine its ability to induce coherent and stable dynamical patterns into the system of chaotic maps interacting with time delay through its directed regulatory links.}
\begin{center}
 ------------------------------------------------
\end{center}

\section{Introduction} \label{Introduction}

The gene regulation is a process of fundamental importance for the functioning and growth of biological cells. Gene regulatory system consists of genes which function collectively by interacting through the transcription factors (activators or repressors), and produce the appropriate proteins in response to cell's needs \cite{uribook}. 

A system of interacting genes can be seen as a \textit{network} \cite{dorogovtsev}, where nodes represent genes and directed links model the activatory and repressory interactions -- such description is termed \textit{Gene Regulatory Network} (GRN) \cite{boccaletti-arenasguilera}. Recent studies revealed various details of GRN's architecture, dynamics and functionality, and allowed applications such as engineering of gene circuits \cite{collins}. GRN of many living organisms are known empirically, both in terms of chemistry of gene interactions and the details of network topology \cite{lee,orr}. A well studied example is bacterium \textit{Escherichia Coli} (E.Coli), where detailed experimental and theoretical investigations revealed all relevant gene interactions \cite{orr,janga}. GRN typically involve different scales of functioning and have \textit{modular} structure, where groups of genes with specific connection patterns preform particular functions \cite{uribook,boccaletti-arenasguilera,orr}.

For modeling the interaction of biological units such as genes, one often uses simple dynamical systems that capture the essence of system's behavior over time \cite{rajesh,jong,areejit,andrecut-coutinho,schuster}. In this way, the dynamics of gene regulations is modeled at various levels \cite{karlebach} and using different mathematical techniques \cite{jong}. The simplest approaches include boolean networks \cite{areejit} and discrete-time maps \cite{andrecut-coutinho}, while more detailed analysis requires 1D \cite{lima} or 2D \cite{schuster} continuous-time ODE, that can also involve a time-delayed action \cite{chunguang}. Analytical studies of GRN models involving small networks revealed the complexity of their dynamical patterns \cite{lima}, with the system of two interacting genes solved in detail \cite{andrecut-coutinho,schuster}. General relationship between structure and function of large GRN was examined, with particular emphasis on their collective dynamics \cite{prill}, information processing \cite{klemm-reliability}, flexibility \cite{areejit}, and functional organization \cite{janga}.

An important question concerning both small and large GRN structures is related to the \textit{stability} of its emergent dynamics: GRN functioning ought to be very robust in order to assure the survival of the cell under various circumstances. How much stability and functionality can a certain network topology provide for a system of dynamical units coupled through its links? General relationship between topology and stability was studied for the case of static \cite{sinha} and growing networks \cite{perotti}. Stability of the GRN dynamics was examined from the Control theory prospective for noisy \cite{jinlindsey} and non-noisy case \cite{chunguang}. As time delay for all interactions is ubiquitous \cite{atay-kurths}, stability and robustness of GRN were also investigated in the case of time-varying delays \cite{rencao}, and asynchronous networks \cite{klemm}. The dynamics at the "edge of chaos" characterized by zero Lyapunov exponents was found in models of gene interaction \cite{dejan}. Some recent studies \cite{shmulevich-nykter} suggest importance of the critical region between the order and chaos in the emergent dynamics of many biological networks. 

Networks of coupled maps are a useful paradigm for designing complex dynamical systems. Networks of maps have been extensively studied on different artificial networks in the context of testing the effects induced by network topology on the emergent phenomena, where typically 1D maps/ODE have been examined \cite{klemm-reliability,sinha}. Chaotic maps have been considered as suitable models that take into account certain features of gene interactions \cite{andrecut-coutinho}, in addition to allowing the study of time delay effects on the emergent behaviors \cite{atay-kurths}. Nonlinear systems were used for modeling brain dynamics \cite{rencao} and temporal fluctuations in gene regulations \cite{klemm}. Coupled maps on empirical networks have been employed as simple tools for examining the robustness of the network topology \cite{li}. Recently, two-dimensional chaotic maps symmetrically coupled with time delay have been studied on artificial scale-free networks. They are found to exhibit a rich variety of the emergent dynamical behaviors \cite{levnajic-tadic,levnajic-teza}, including strange non-chaotic attractors \cite{feudel}. 

The purpose of the present work is somewhat different: we focus on a realistic Gene Regulatory Network of bacterium E.Coli, and study its dynamical stability using a different class of nonlinear dynamical systems as our tool. In particular, we employ two-dimensional chaotic maps associated with the nodes, while varying the coupling types and strengths. We determine the conditions under which the gene regulatory architecture maintains its dynamical stability and describe the pathways towards the stable collective dynamical states, which contrasts the strongly chaotic dynamics exhibited by the isolated nodes. We describe the emergent states of the network by several quantitative measures.

For the nonlinear chaotic map associated with each node we use 2D Chirikov standard map \cite{ll}, which is well understood and shares some {\it formal} properties (specifically the dimensionality of the phase space) with the actual gene dynamics, given at each node by the concentrations of protein and \textit{m}RNA. It contains built-in periodicity in the angle variable, while being unbounded in the action variable. We stress that this choice is primarily motivated by a tunable chaotic behavior of the isolated standard map, which allows an easy quantification of the emergent non-chaotic dynamics that is a clear consequence of the network interaction among the maps. Our system is \textit{dissipative}, despite involving conservative standard maps as units. Dissipativeness is a general feature of biological dynamics, which in our model arises as a consequence of the particular coupling form that we chose, in addition to the network's directedness.\\[0.01cm]

\begin{figure}[!hbt]  \begin{center}
\includegraphics[height=2.7in,width=3.4in]{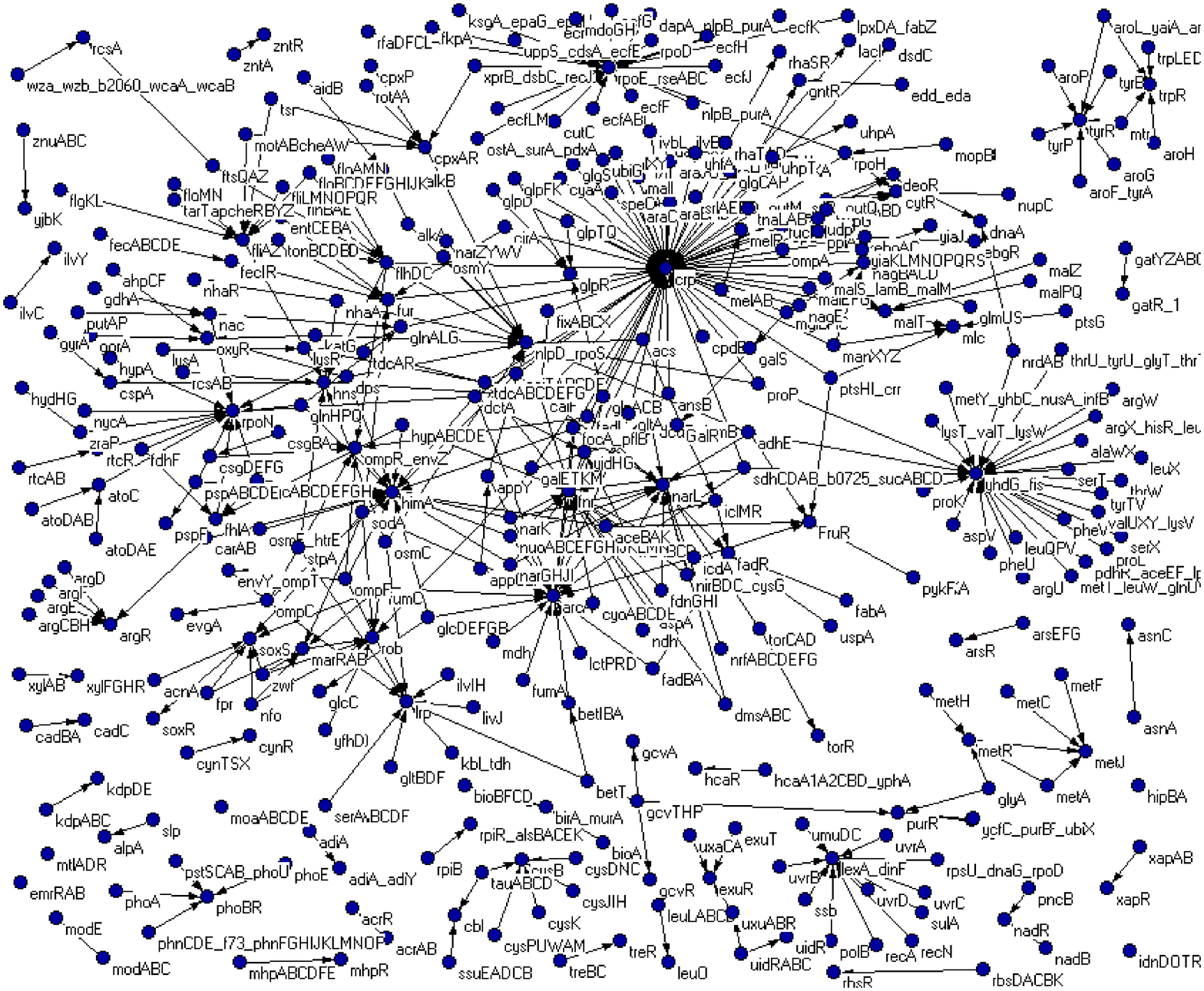}
\caption{Directed E.Coli's GRN with $N=423$ genes marked by their biological names, according to the data from \cite{orr,uriweb}. Self-loops as well as the difference between
 activatory and repressory links are omitted.} \label{ecoli-CCM-figure-1}
\end{center} \end{figure}

\n \textbf{The Network.} We consider the version of E.Coli's GRN shown in Fig.\,\ref{ecoli-CCM-figure-1}, which was
empirically determined in 2003 \cite{orr} (data available in Ref.\cite{uriweb}). Since we are interested in the induced collective dynamics, we consider only the largest connected component, which is composed of $N=328$ nodes/genes (shown in Figs.\,\ref{ecoli-CCM-figure-6}\,\&\,\ref{ecoli-CCM-figure-8}). Smaller components have their own mutually unrelated dynamics. Compared to our previous works where the coupled maps were studied on non-directed networks \cite{levnajic-tadic,levnajic-teza}, the focus here is on the effects that the directedness of links with time-delayed interactions have on the emergent collective dynamics and its stability.\\[0.01cm]

The paper is organized as follows: in Section \ref{Chaotic Maps on Directed E.Coli GRN}. we introduce the system of chaotic maps coupled through the directed links of E.Coli's GRN and study briefly the case of SUM-rule coupling. Our main results are given in Section \ref{Chaotic Maps on E.Coli GRN with Spreading Coupling}. where we study the Spreading coupling between the maps and analyze in detail the collective dynamics in two characteristic regions of the  coupling strengths. A short summary and the discussion of the results are given in Section \ref{Conclusions}.


\section{Chaotic Maps on Directed E.Coli GRN} \label{Chaotic Maps on Directed E.Coli GRN}

In this Section we define the directed network of chaotic 2D maps and introduce statistical and stability measures that will be used to characterize its emergent dynamics. We also show the main results for the case of SUM-rule coupling, which are compared to the results for non-directed case \cite{levnajic-tadic}.\\[0.01cm]

\n \textbf{The Network of Chaotic Maps.} We consider the largest connected component of the directed E.Coli's GRN, shown in Figs.\,\ref{ecoli-CCM-figure-6}\,\&\,\ref{ecoli-CCM-figure-8}, and assign to each node $[i]=1,\hdots N$ a Chirikov standard map which consists of two mutually coupled (action-angle) variables:
\begin{equation} \left( \begin{array}{c}
x[i]' \\
y[i]'
\end{array} \right) = \left(
\begin{array}{l} 
  x[i] + y[i] + \e \sin (2 \pi x[i])  \;\;  \mbox{mod} \; 1  \\
  y[i] + \e \sin (2 \pi x[i])
\end{array} \right) \label{oursm} \end{equation}
representing a discrete version of the kicked rotor \cite{ll}. The chaotic parameter $\e$ is set to $\e=0.9$, thus assuring a strongly chaotic behavior of the isolated units. Each node is described by its dynamical state in two-dimensional phase space, represented by a point $(x[i]_t,y[i]_t)$ at a discrete time $t$. The maps are coupled along the directed network links, with a coupling term that involves a one-step time delay in the coupled angle coordinates $x$ between the network neighbors:
\begin{equation} 
\begin{tabular}{l}
$\left(\begin{array}{l}
x[i]_{t+1} \\
y[i]_{t+1}
\end{array}\right)
=(1- \mu) 
\left(\begin{array}{l}
x[i]_t' \\
y[i]_t'
\end{array}\right)+$ 
      \\ 
$\;\;\;\;\;\;\;\;\;\;\;\;\;\;\;\;\;\;\;\;
M(\mu,{k_i^{in}})
\left(\begin{array}{c}
\sum_{j \in {\mathcal K_i}} (x[j]_t - x[i]_t') \\ 
0 
\end{array}\right)$
\end{tabular}
\label{directed+conn-equation} \end{equation}
$M(\mu,k_{i}^{in})$ defines the type of interaction between the nodes, which always depends on the coupling strength $\mu$. Note that in a directed network ${\mathcal K_i}$ denotes only the neighbors having links \textit{towards} the node $[i]$, and accordingly, the interaction term $M(\mu,k_{i}^{in})$ involves only the node's in-degree $k_{i}^{in}$. Some nodes may not have incoming links ($k_i^{in}=0$) and have thus vanishing inputs from other nodes. However, their motion is eventually affected due to $(1-\mu)$ term, which inhibits the chaotic diffusion \cite{levnajic-teza}. Time delay is realized through the fact that the coupling term contains the node's updated value ($x[i]_t'$), while the neighboring nodes' values ($x[j]_t$) are not updated (for an update we here intend the action of the standard map denoted as $' $ and defined in Eq.\,(\ref{oursm})).

In this Section we briefly discuss the case with the coupling form
\begin{equation}
M(\mu,k_i^{in}) = \mu k_i^{in} \label{eq-sumrule}
\end{equation}
which resembles the SUM logic-gate GRN model: the input that a gene receives is the average of the inputs coming from all the neighbors \cite{jinlindsey} (a different type of coupling is discussed in Section \ref{Chaotic Maps on E.Coli GRN with Spreading Coupling}.). It is important to stress that the neighboring nodes are coupled via their angular variables $x$ only, while the dynamics of the action coordinate $y$ is affected via standard map update Eq.\,(\ref{oursm}), and by the prefactor $(1-\mu)$ in Eq.\,(\ref{directed+conn-equation}).\\[0.01cm]

\n \textbf{Types of Orbits and Statistical Measures Employed.} In the simulations with the dynamics defined by Eqs.\,(\ref{oursm}-\ref{directed+conn-equation}), we start with a random selection of initial conditions from the interval $(x,y) \in [0,1] \times [-1,1]$ for each GRN node, and a fixed coupling $\mu$ along the network links. The transients are taken as $t_0=10^{5}$ iterations. Two types of the emergent orbits are considered:
\begin{itemize}
\item orbit of the individual nodes:  $(x[i]_t,y[i]_t)_{t>t_0}, \;\; [i]=1,\hdots, N$.
\item time-averaged orbit for a given node: $(\bar{x}[i],\bar{y}[i]) =  \lim_{t \rightarrow \infty} \frac{1}{t-t_0}\sum_{k=t_0}^{t} (x[i]_k,y[i]_k)$ which reduces an entire orbit of a given node into a single phase space point. 
It qualitatively captures its motion after transients (or during the time-interval $[t_0,t]$ if $t<\infty$). 
\end{itemize}
As mentioned above, we are testing the ability of the examined network structure to regularize the motion of the chaotic maps due to their mutual interactions along the directed links. The system's emergent dynamics will be characterized using following tools:\\[0.08cm]
$\bullet$ \textit{Periodic orbits} for single nodes are defined to have a periodicity $\pi = t_1 - t_0$ if for some $t_1>t_0$ we have
\[ \Bigg| \dfrac{x[i]_{t_0} - x[i]_{t_1}}{x[i]_{t_0}} \Bigg|  < \delta \;\;\; \mbox{and} \;\;\;  \Bigg| \dfrac{y[i]_{t_0} - y[i]_{t_1}}{y[i]_{t_0}}\Bigg|  < \delta \] 
Here we set  $\delta=10^{-4}$. Appearance of stable periodic orbits indicates a presence of regularity in the collective dynamics, in a clear contrast with the chaotic behavior of the isolated standard map.\\[0.08cm]
$\bullet$ \textit{Finite-time Maximal Lyapunov Exponent} (FTMLE) for a point $\x_0$ is defined as the maximal initial divergence rate between $\x_0$ and the trajectories starting in its neighborhood ${\mathcal N}$:
\[ \Lambda_{max}^t (\x_0) = \max_{\x \in {\mathcal N}} \; \left\lbrace \mbox{initial slope}  \left[ \frac{1}{\tau}\ln \frac{d(U_\tau \x,U_\tau \x_0)}{d(\x,\x_0)} \right]\right\rbrace \] 
where the operator $U_\tau$ denotes the time-evolution according to Eqs.\,(\ref{oursm}-\ref{directed+conn-equation}) until the time $\tau$. The actual $\tau$-value used in this expression is determined for each particular orbit, which can be identified as either strongly or weakly chaotic, or otherwise regular. We first determine the character of the orbit by computing an approximate FTMLE using $\tau=10^3$ iterations, and only then set the $\tau$-value used to compute the real FTMLE. Typically, $\tau$ ranges from 20 iterations for strongly chaotic to 200 iterations for stable orbits. For a given emergent orbit $\Lambda_{max}^t$-s are averaged for many initial points $\x_0$ belonging to it, and $\lambda_{max}^t = \langle \Lambda_{max}^t (\x_0) \rangle_{\x_0}$ is considered the characteristic FTMLE for this orbit, quantifying its stability. In opposition to the standard Lyapunov exponents typically used with isolated chaotic systems \cite{levnajic-teza}, FTMLE are a suitable measure for single nodes attached to the rest of the network, thus giving a node-by-node characterization of the network stability.\\[0.08cm]
$\bullet$ \textit{Return-times to Phase space Partitions} represent an interesting measure of the non-ergodicity of the dynamics \cite{levnajic-tadic}, and thus provides a suitable way of quantifying the collective effects. We monitor the time intervals between successive visits of an orbit to a phase space partition element. As discussed in Section \ref{Chaotic Maps on E.Coli GRN with Spreading Coupling}., for this purpose we consider $10^5$ equal elements along the $x$-coordinate, while the cells remain unlimited along the $y$-coordinate.\\[-0.1cm]

\n \textbf{The Case of SUM-rule Coupling.} We find that the GRN network with the SUM-rule coupling between the units given by Eq.\,(\ref{directed+conn-equation}-\ref{eq-sumrule}) exhibits periodic orbits with different periodicities for 
\begin{figure}[!hbt]  \begin{center}
   \includegraphics[height=4.2in,width=2.8in]{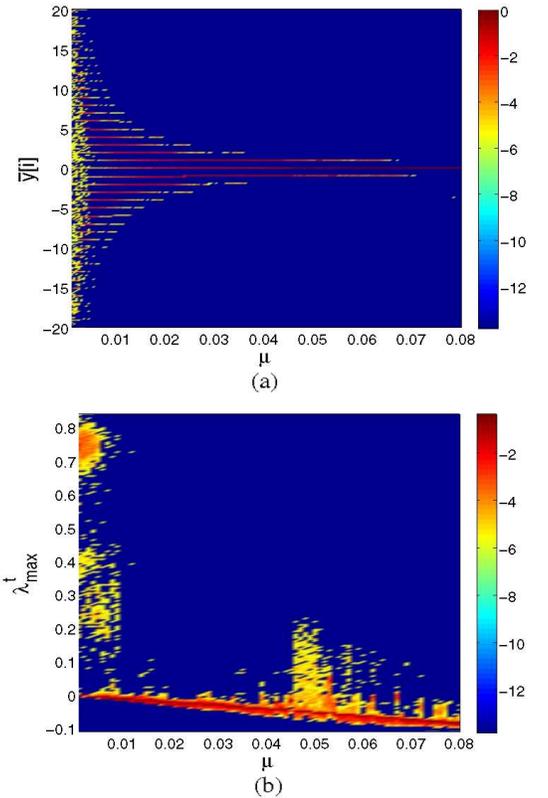} 
\caption{2D histogram of $\bar{y}$-values in (a), and $\lambda^{t}_{max}$-values in (b). The color-map shows the fraction of nodes having a certain value of $\bar{y}$ or $\lambda^{t}_{max}$ as function of the coupling $\mu$. Log-scale is used for the color-map, with the lowest (reference) value taken as $10^{-14}$, corresponding to the deep blue color.} \label{ecoli-CCM-figure-2}
\end{center} \end{figure}  
any non-zero $\mu$-value, as shown in Fig.\,\ref{ecoli-CCM-figure-2}. In addition, self-organized non-periodic orbits are found at some nodes for the coupling strengths around $\mu \sim 0.05$. To compare with the case of non-directed (symmetrical) network from \cite{levnajic-tadic}, we study the clustering and stability of the emergent orbits. A histogram of $\bar{y}[i]$-values as function of $\mu$ is reported in Fig.\,\ref{ecoli-CCM-figure-2}a, showing groups of nodes with clustered orbits. In Fig.\,\ref{ecoli-CCM-figure-2}b the histogram for the corresponding  FTMLE values $\lambda_{max}^t$ is shown. Both histograms are obtained for a single set of initial conditions at every $\mu$-value and averaged over all the network nodes.

These results are qualitatively similar to the case of non-directed networks: the system exhibits clusters of stable periodic orbits for any non-zero coupling, along with weakly chaotic orbits (small positive $\lambda_{max}^t \sim 0.1$) around $\mu \sim 0.05$ for a some small fraction of nodes. Strongly chaotic orbits similar to the isolated standard map's ones are occurring for very small $\mu$-values, and disappear for larger $\mu$-values. The number of clusters decreases and eventually shrinks to one. In analogy with the non-directed coupling, three dynamical regions can be distinguished: strongly chaotic ($\mu \lesssim 0.01$), periodic ($0.01 \lesssim \mu \lesssim 0.04$), and weakly chaotic with self-organized orbits ($\mu \gtrsim 0.04$). The network dynamics for large coupling strengths (beyond the shown interval) remains stable with all orbits being periodic. We conclude that with the appropriate coupling form -- SUM-rule with the normalized inputs at each node -- E.Coli's directed GRN is generally able to regularize the behavior of the coupled chaotic maps in a wide range of coupling strengths.


\section{Chaotic Maps on E.Coli GRN with Spreading Coupling} \label{Chaotic Maps on E.Coli GRN with Spreading Coupling}

We now modify our network of coupled maps by employing the Spreading coupling form. In Eq.\,(\ref{directed+conn-equation}) we set:
\begin{equation}   M(\mu,k_i^{in}) = \mu = const. \label{eq-spreadingrule} \end{equation}
We further test the potentials of the directed E.Coli's GRN to stabilize the behavior of chaotic standard map associated with its nodes, by investigating the emergent dynamics of Eqs.\,(\ref{directed+conn-equation}\,\&\,\ref{eq-spreadingrule}) using the tools and the approach introduced in the previous Section. In contrast to the SUM-rule coupling studied above, in this case the system is expected to be more sensitive to the chaotic behavior, given that better connected nodes receive larger total inputs compared to the nodes with fewer links. With this type of coupling the dissipativeness becomes unevenly distributed over nodes, with better linked nodes undergoing stronger dissipative effects. As it will be shown below, this inhomogeneity of the dissipation dramatically affects the collective dynamics. 

\begin{figure}[!hbt]
\begin{center}
\includegraphics[height=2.4in,width=3.2in]{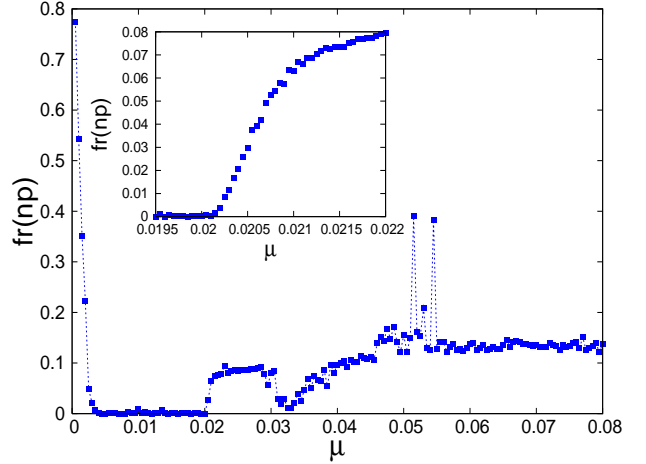}
\caption{Fraction of non-periodic orbits as function of the coupling strength $\mu$, for all nodes and averaged over many initial conditions. Inset: zoom to the dynamical transition region around $\mu \sim 0.022$.}  \label{ecoli-CCM-figure-3}
\end{center}
\end{figure}

In Fig.\,\ref{ecoli-CCM-figure-3} we report the fraction of non-periodic orbits as function of $\mu$ for all network nodes, averaged over many initial conditions. The profile of the curve at small $\mu$-values resembles the one for the non-directed network and the SUM-rule cases: for all initial conditions, after a quick initial transient all nodes stabilize into different periodic orbits. However, with the increase of the coupling strength a
fraction of nodes is found around $\mu \sim 0.02$, whose emergent orbits remain non-periodic for all initial conditions (inset in Fig.\,\ref{ecoli-CCM-figure-3}). Generally, with further increase of $\mu$ the fraction of nodes with non-periodic orbits fluctuates, but overall slowly increases. Properties of these nodes and their orbits will be studied in detail below.

\begin{figure}[!hbt]
\begin{center}
   \includegraphics[height=4.2in,width=2.8in]{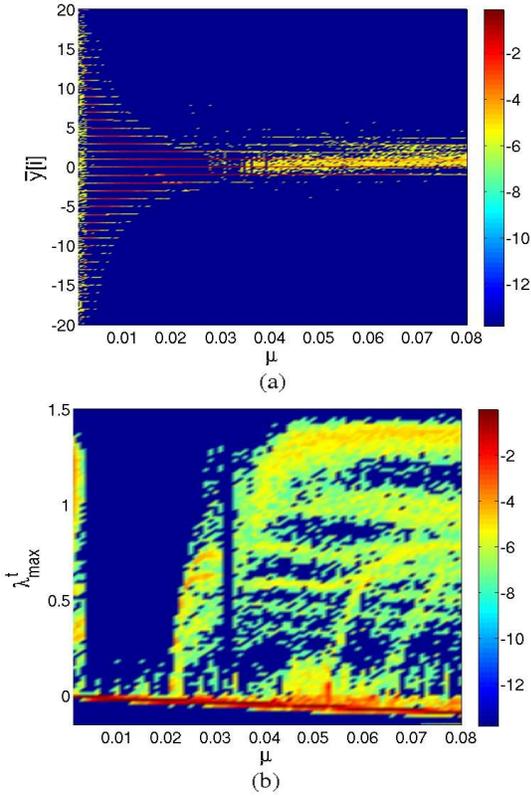} 
  \caption{2D histogram of $\bar{y}[i]$-values in (a) and $\lambda_{max}^{t}$-values in (b) for directed network dynamics Eqs.\,(\ref{directed+conn-equation}\,\&\,\ref{eq-spreadingrule}), computed for a single initial 
     condition and for all nodes. The log-scale color-map is as explained in Fig.\,\ref{ecoli-CCM-figure-2}.} 
  \label{ecoli-CCM-figure-4}
\end{center}
\end{figure}

The 2D histogram of $\bar{y}[i]$-values, shown in Fig.\,\ref{ecoli-CCM-figure-4}a, is reflecting different dynamical regions with the occurrence of the non-periodic orbits, in agreement with Fig.\,\ref{ecoli-CCM-figure-3}. 
For small $\mu$-values we find a fully clustered organization of nodes with periodic orbits only, whereas for $\mu \gtrsim 0.02$ only a few separate clusters of orbits can be identified. Contrary to the SUM-rule case, here we find that for $\mu \gtrsim  0.08$ the network displays mainly non-clustered orbits, co-existing with a few remaining clusters. The situation is also reflected in the 2D histogram of the FTMLE $\lambda_{max}^{t}$, shown in Fig.\,\ref{ecoli-CCM-figure-4}b. For small non-zero couplings $\mu$, non-positive $\lambda_{max}^{t}$ can be found at all nodes. However, destabilization of nodes occurs for $\mu \gtrsim 0.02$, where we find a spectrum of positive $\lambda_{max}^{t}$-values. In this region of coupling strengths $\lambda_{max}^{t} \gtrsim 1$ is found at some nodes, indicating their strongly chaotic behavior, similarly to the isolated standard map. Apart from exhibiting the unstable dynamics with positive FTMLE at the majority of nodes, the 2D histogram of $\lambda_{max}^{t}$-values for $\mu>0.022$ exhibits a characteristic pattern, as visible in Fig.\,\ref{ecoli-CCM-figure-4}b, that includes a spectrum of small FTMLE $\lambda_{max}^{t} \gtrsim 0$. In addition, this patterns suggests that the weakly chaotic nodes having $\lambda_{max}^{t} \gtrsim 0$ occur as a large population of the network nodes, in opposition to a much smaller fraction of nodes with $\lambda_{max}^{t} \gtrsim 1$. In what follows we study in detail the emergent dynamics at all nodes on the network for two particular coupling strengths: at the onset of destabilization for $\mu \sim 0.022$, and within the unstable region at $\mu=0.05$.\\[0.01cm]

\begin{figure}[!hbt] \begin{center}
   \includegraphics[height=4.2in,width=2.8in]{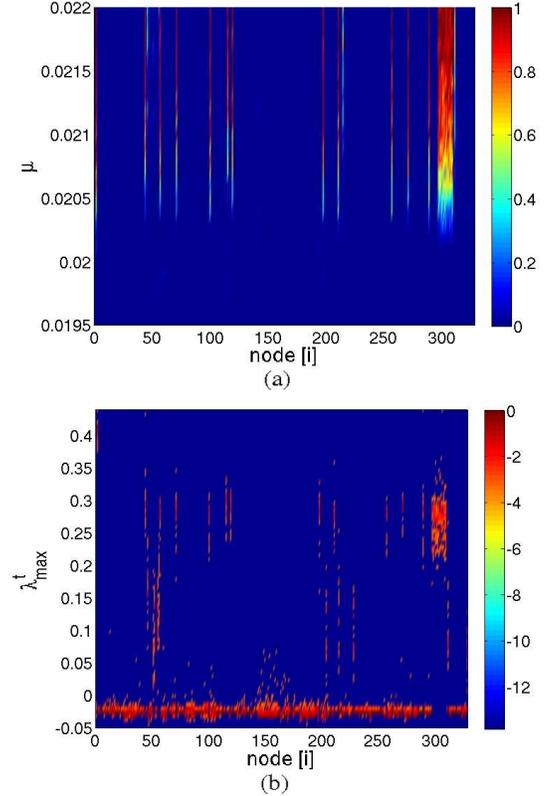} 
  \caption{For each network node $[i]$ the color-map shows the fraction of initial conditions that: (a) lead to non-periodic orbits as a function of $\mu$ in the vicinity of  the threshold, and (b) lead to a 
given $\lambda_{max}^{t}$-exponent for the threshold coupling $\mu=0.022$.}   \label{ecoli-CCM-figure-5}
\end{center} \end{figure}

\n \textbf{Dynamics at the Onset of Destabilization $\mu \sim 0.022$.} The collective dynamics of the directed E.Coli's GRN described by Eq.\,(\ref{directed+conn-equation}) with the Spreading coupling rule Eq.\,(\ref{eq-spreadingrule}) becomes partially unstable for $\mu>0.02$, exhibiting a mixture of stable and weakly chaotic orbits. As shown in the inset in Fig.\,\ref{ecoli-CCM-figure-3}, the destabilization occurs gradually, resembling a continuous phase transition. For the coupling strengths close to the threshold $\mu \sim 0.022$ we examine the fraction of non-periodic orbits at each node for many initial conditions. The results are shown in Fig.\,\ref{ecoli-CCM-figure-5}a. 
We can identify the nodes that gradually lose periodic orbits as the intensity of coupling increases beyond $\mu=0.022$. These nodes seem to undergo the process almost simultaneously, while the rest of the network appears to be unaffected. In order to further analyze the nature of the non-periodic orbits, we compute and show in Fig.\,\ref{ecoli-CCM-figure-5}b the histogram of the $\lambda_{max}^{t}$-values for each network node separately, at this particular coupling strength of $\mu=0.022$. Indeed, the nodes that lose periodic orbits display $\lambda_{max}^{t}>0$. Positive $\lambda_{max}^{t}$ are smaller than 1, indicating weakly chaotic effects at the unstable nodes. It is striking feature of this directed network that the instability develops on a specific group of nodes, rather than over the whole network. Moreover, this instability remains confined to the initially destabilized sub-network even after the transition at $\mu=0.022$. Whereas, the rest of the network manages to maintain its regular stable dynamics.

\begin{figure}[!ht] \begin{center}
\includegraphics[height=2.67in,width=3.2in]{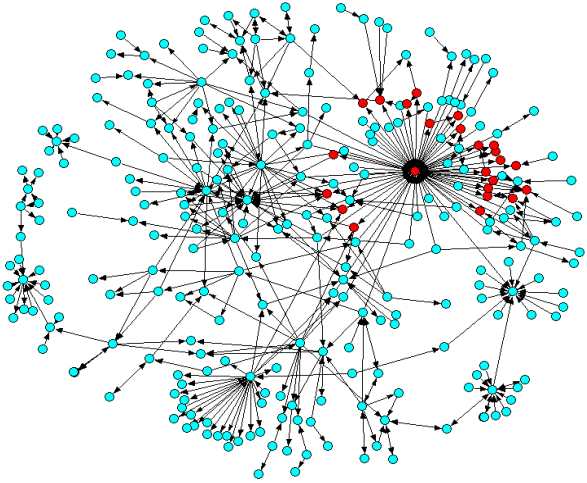}
\caption{Largest connected component of directed E.Coli GRN. Nodes exhibiting non-periodic orbits at $\mu=0.022$ are marked by red, and those with periodic orbits by blue color.} 
     \label{ecoli-CCM-figure-6}
\end{center} \end{figure}

In Fig.\,\ref{ecoli-CCM-figure-6} we show the largest connected component of the directed E.Coli's GRN indicating the node stability: nodes with periodic orbits at $\mu=0.022$ are shown in blue and non-periodic ones in red color. The unstable nodes are mutually connected and form a sub-network composed of 24 nodes, and include the hub-node (a list of biological names of these unstable nodes/genes is given in Ref.\cite{0022names}). 
The unstable sub-network has both incoming and outgoing links with the rest of the network. This suggests a non-trivial interplay between the collective dynamics and the network architecture, responsible for keeping the unstable dynamics co-existing with the stable one within the same network. We run the simulations for longer than $10^6$ iterations to confirm the stationarity of these results.

\begin{figure*}[!ht] \begin{center}
\includegraphics[height=4.45in,width=5.2in]{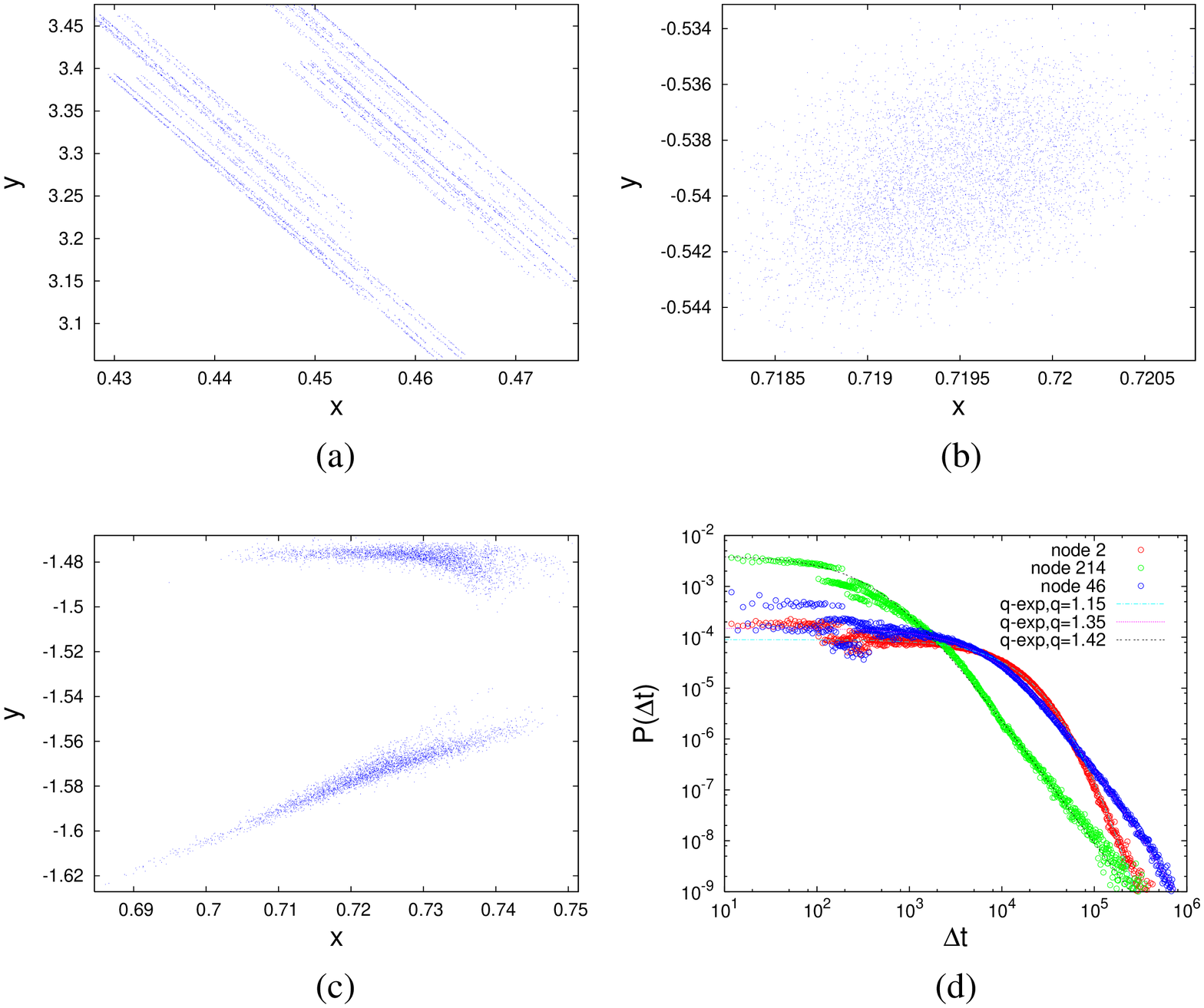}
\caption{Examples of typical orbits for unstable single nodes at $\mu=0.022$: (a) the attractor of the hub-node 2 (gene biological name "acrAB"), (b) an orbit of the node 214 (gene "kbl-tdh"), and (c) an orbit of the node 46 (gene "aspV"). Only representative parts of the orbits are shown. (d) distributions of return times to phase space partition for these single-node orbits, fitted with the $q$-exponential expression Eq.\,(\ref{eq-qexp}), with the respective $q$-values listed. Phase space is partitioned into $10^5$ equally spaced cells in $x$-coordinate ($y$-coordinate unbounded).}  \label{ecoli-CCM-figure-7}
\end{center}\end{figure*}

In Fig.\,\ref{ecoli-CCM-figure-7}a,\,b\,\&\,c three typical attractors/orbits appearing on single unstable nodes at $\mu=0.022$ are shown, exhibiting different structural patterns within reduced areas of the phase space. The hub-node always displays the strange attractor from Fig.\,\ref{ecoli-CCM-figure-7}a with a fractal structure, regardless of the initial conditions ($\lambda_{max}^{t} \simeq 0.4$, fractal dimension $d_f \simeq 1.4$). We examine the statistical properties of these single-node orbits by computing their return time distributions with respect to the phase space partitions, defined above. The obtained distributions are fitted with $q$-exponential function defined as \cite{levnajic-teza}:
\begin{equation}   e_q(x) = B_q \left( 1 - (1-q)\frac{x}{x_q} \right)^{\frac{1}{1-q}}  \label{eq-qexp} \end{equation} 
The results for all three orbits are shown in Fig.\,\ref{ecoli-CCM-figure-7}d. The distributions indicate long-time correlations in the return times, compatible with the power-law tail having $q>1$. The trajectories explore different parts of the phase space non-uniformly, in a clear opposition to the exponential distribution (for $q=1$), which characterizes full ergodicity of the chaotic dynamics exhibited by the isolated nodes. Occurrence of the power-law tails of the distributions suggests the presence of self-organization effects, arising as a consequence of the interaction among the nodes, which leads to the emergent non-periodic orbits with $\lambda \gtrsim 0$.

In a sharp contrast with what was observed in the previous Section, the directed network with the dynamics given by Eqs.\,(\ref{directed+conn-equation}\,\&\,\ref{eq-spreadingrule}) remains destabilized above the coupling intensity of  $\mu=0.022$. Also, contrary to the uniformity of the emergent states in the case of the symmetrically coupled maps, in the present case we find that several patterns of different dynamical nature can be simultaneously present in different parts of the network.\\[0.01cm]

\begin{figure}[!hbt] \begin{center}
\includegraphics[height=2.67in,width=3.2in]{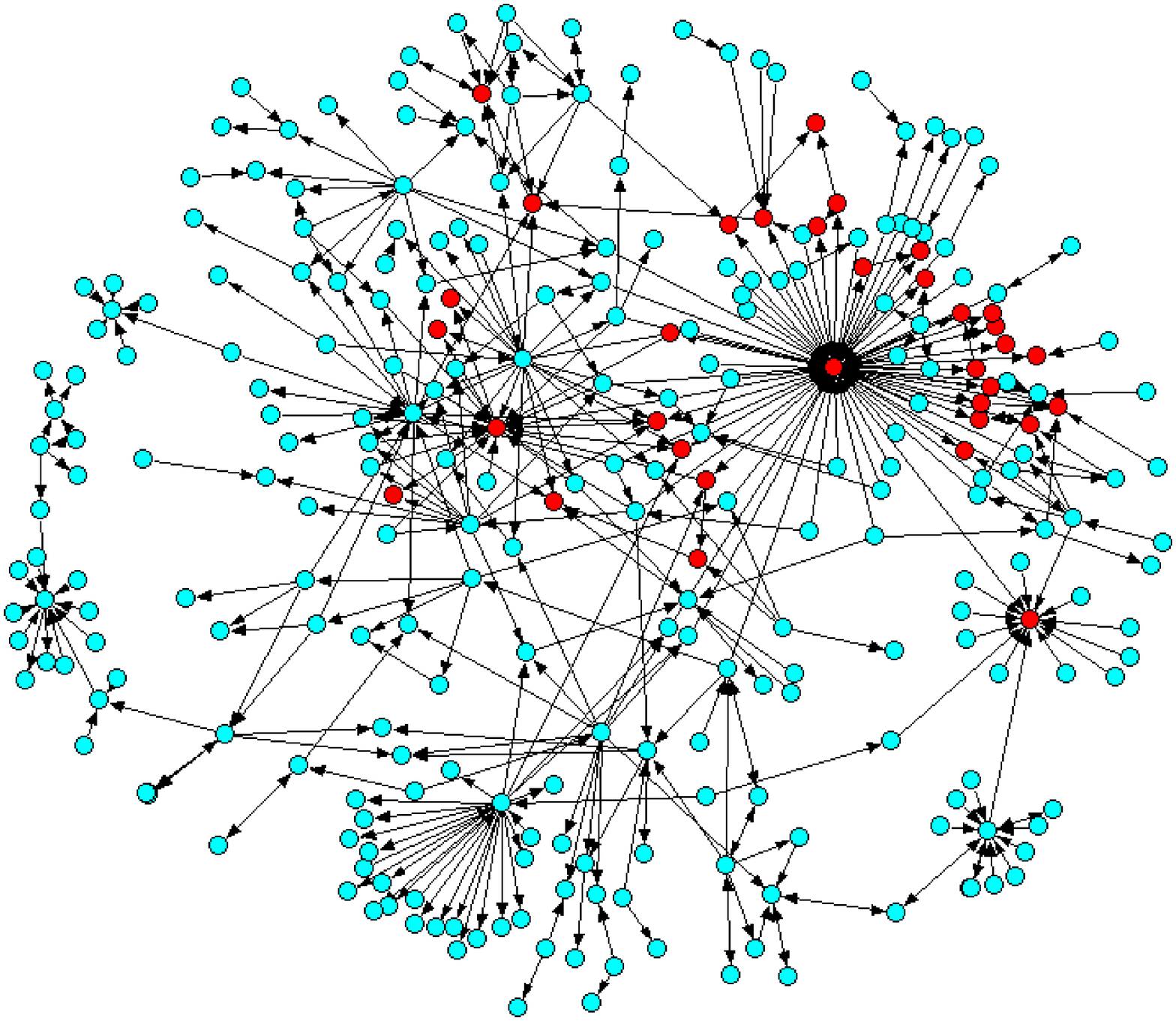}
\caption{Coupled chaotic maps on the directed E.Coli GRN for $\mu=0.05$: non-periodic nodes are localized in the red sub-network, while the rest of nodes (blue) manage to attain the periodic orbits.} 
     \label{ecoli-CCM-figure-8}
\end{center} \end{figure}

\n \textbf{Collective Dynamics inside Chaotic Region at $\mu=0.05$.}  In the region of coupling strength $\mu \gtrsim 0.022$ the GRN nodes are exhibiting three types of behavior: stable periodic orbits, weakly chaotic orbits, and strongly chaotic orbits (cf. Figs.\,\ref{ecoli-CCM-figure-3}\,\&\,\ref{ecoli-CCM-figure-4}). In Fig.\,\ref{ecoli-CCM-figure-8} we show the state of the network at $\mu=0.05$, where the non-periodic nodes are indicated by red color. The unstable behavior is again localized to a connected sub-network, which is integrated within the rest of the network. The unstable sub-network comprises 10\% of the total number of nodes and includes the unstable sub-network found in case of $\mu=0.022$ (with additional 10 nodes listed in Ref.\cite{005names}). 
Despite stronger coupling, the network still manages to contain its dynamically unstable part within a small sub-network, allowing for regular stable behavior of the rest of the nodes.

\begin{figure}[!hbt] \begin{center}
   \includegraphics[height=4.3in,width=2.8in]{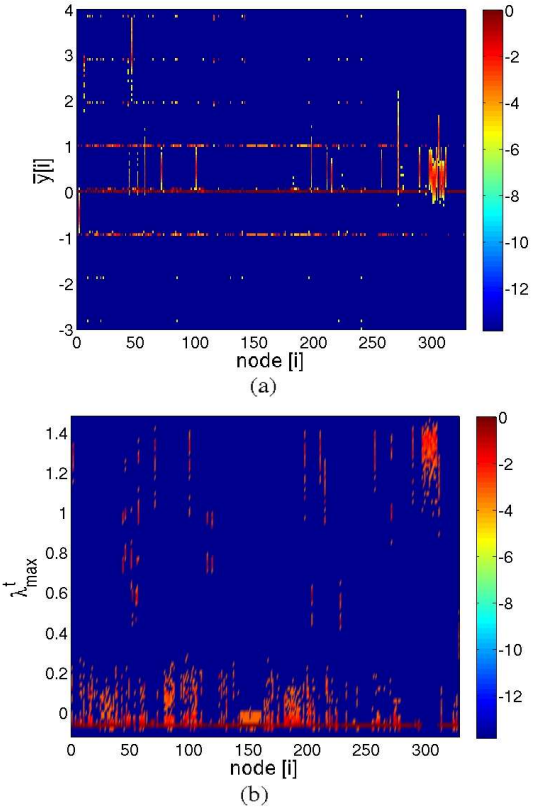} 
\caption{For each node $[i]$ the color-map shows the fraction of initial conditions for $\mu=0.05$ leading to: a given value of $\lambda_{max}^t$-exponent in (a) and a given $\bar{y}[i]$-value in (b). Log-scale is used as with Fig.\,\ref{ecoli-CCM-figure-2}.} \label{ecoli-CCM-figure-9}
\end{center} \end{figure}

In Fig.\,\ref{ecoli-CCM-figure-9}a we show 2D histogram of $\bar{y}[i]$-values node-by-node at $\mu=0.05$. While most of the nodes' orbits are clustered into three main clusters for all initial conditions (horizontal lines in this plot), some other nodes have a wide range of possible $\bar{y}[i]$-values away from a clear clustering pattern. For comparison, in Fig.\,\ref{ecoli-CCM-figure-9}b we also show the histogram of $\lambda_{max}^t$-values for each node on the network and the fixed coupling strength $\mu=0.05$.
As the Fig.\,\ref{ecoli-CCM-figure-9}b shows, the majority of nodes appear to have stable dynamics ($\lambda_{max}^t<0$). For the unstable nodes we find the values of $\lambda_{max}^t$ to vary from $\lambda_{max}^t \sim 0.1$ (also present at the instability threshold $\mu=0.022$), to a strong chaos with $\lambda_{max}^t \gtrsim 1$. We also find that some of these nodes alternate between stable and weakly unstable dynamics. In our coupled maps dynamics on directed GRN, each node attains a certain spectrum of "roles" in the network emergent behavior, depending on the particular choice of the initial conditions and the connection patterns, thus implying the roles of other nodes connected to it.

\begin{figure*}[!hbt] \begin{center}
\includegraphics[height=4.45in,width=5.2in]{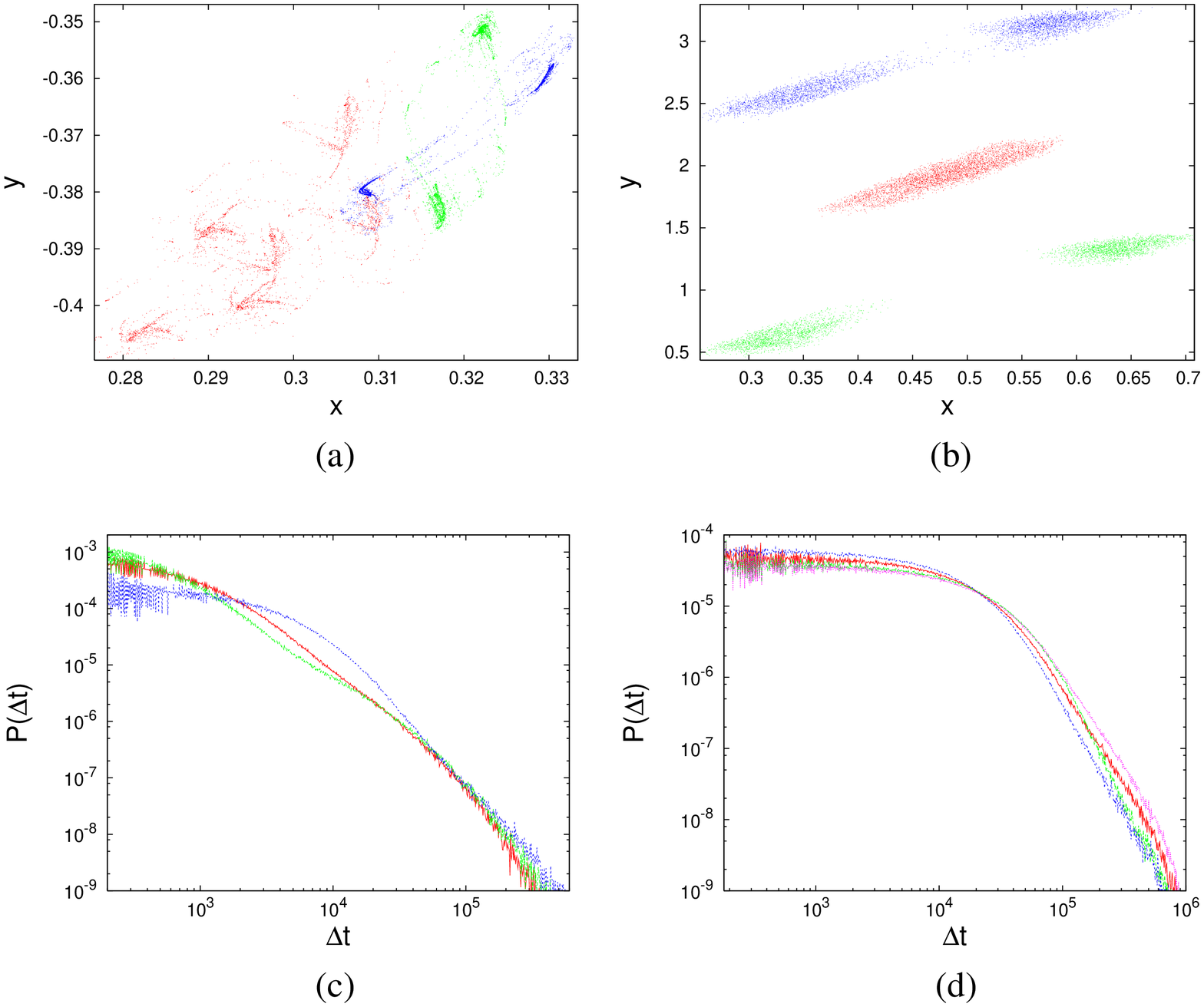}
\caption{Examples of three emergent orbits (marked by different colors) for single nodes at $\mu=0.05$: the node 86 (gene "zraP") in (a), and the node 119 (gene "fdhF") in (b). The distributions of return times to phase space partitions are shown for the orbits from (a) in (c), and for the orbits from (b) in (d). Phase space partitioning is done identically to the case from the Fig.\,\ref{ecoli-CCM-figure-7}.}  \label{ecoli-CCM-figure-10}
\end{center}  \end{figure*}

Contrary to the case of $\mu=0.022$ shown in Fig.\,\ref{ecoli-CCM-figure-7}, at the strong coupling the hub node does not display any specific attractor, but it shows strongly chaotic motion regardless of the initial conditions. However, the behavior of some other non-periodic nodes shows a variety of patterns that are structurally and statistically robust to the initial conditions. We illustrate this in Fig.\,\ref{ecoli-CCM-figure-10}, where three typical emergent orbits of two selected unstable nodes are shown, along with the corresponding distributions of the return times. The key geometrical and statistical properties of the orbits seem to persist for varying initial conditions and are related with the observed fluctuations in $\bar{y}[i]$-values. The shapes of return times distributions for these orbits are shown in Fig.\,\ref{ecoli-CCM-figure-10}c\,\&\,d. Again, these distributions can be fitted with the expression Eq.\,(\ref{eq-qexp}) with $q>1$, indicating an self-organized collective dynamics, away from the fully chaotic dynamics of the isolated maps.

The emergent dynamics of our coupled chaotic maps system possesses a certain {\it flexibility} with respect to the initial conditions (which can be seen as the environmental inputs to the cellular GRN). It is marked by the finite range of possible $\bar{y}[i]$-values and $\lambda_{max}^t$-values. On the other hand, the statistical properties of the dynamics, i.e., the return-time distributions of the angular variable, for particular unstable nodes/genes, appear to be quite robust to variations of the initial conditions. Features of this type might be related to the biological origin of the studied network: the network's architecture provides certain level of functional adaptability and robustness for operation of its units - genes, even when the inherently chaotic dynamics at its units is present.


\section{Discussion and Conclusions} \label{Conclusions}

We have studied the emergent dynamics and stability of two-dimensional chaotic standard maps coupled with time delay along the directed regulatory links of the largest connected component of Escherichia Coli's Gene Regulatory Network. The coupling with SUM-rule and Spreading-rule between the connected units have been considered, and the emergent dynamical properties were investigated through non-periodic orbits, and by using Finite-time Maximal Lyapunov Exponents $\lambda_{max}^t$ and distributions of return-times to the phase space partitions. We have demonstrated that the examined empirical network structure is able to induce a coherent collective dynamics of the coupled chaotic maps. The emergent behavior appears to depend primarily on the network structure, but also on the type and the strength of coupling.

Our numerical results suggest that different mechanisms are in a combined manner contributing to the observed collective dynamical behavior of the coupled maps on network:
\begin{itemize}
\item {\it Self-organization} among dynamically coupled units occurs over time, where the state of each unit adjusts to the states of its neighbors in various ways, depending on the couplings and the local dynamics of its two variables, without any external interaction involved.
\item {\it Anomalous diffusion} in our coupled map network is demonstrated for the trajectories of some nodes, which are shown to selectively fill the phase space.
\item {\it Non-ergodicity}, related with the anomalous diffusion and the self-organized behavior of the coupled maps, is a general property of complex dynamical systems. In the present case it might be caused by the underlying network structure, combined with the dissipatively coupled maps on it.
\item {\it Synchronization} occurring between different nodes is documented by the appearance of the clusters of nodes with a well defined $\bar{y}[i]$-values (cf. Figs.\,\ref{ecoli-CCM-figure-4}\,\&\,\ref{ecoli-CCM-figure-9}).  Mutually synchronized nodes tend to form a characteristic pattern over the network which we have not studied for this case (results for non-directed networks can be found in \cite{levnajic-tadic,levnajic-teza}).
\item {\it Phenomena at the edge-of-chaos} or a dynamical phase-transition are often reported (see e.g. \cite{dejan,shmulevich-nykter}) in the extended dynamical systems when the Lyapunov exponents vanish. Our results also suggest that the attractors with vanishingly small finite-time Lyapunov exponents $\lambda_{max}^t$ might play an important role in the occurrence and the structure of the collective dynamical states on the studied GRN.
\end{itemize}

The structural properties of the network are closely related with the observed dynamical features and their stability. The flexibility of the emergent orbits to the initial conditions indicates the network's ability to efficiently respond to the environmental inputs by adapting dynamics of its units accordingly. Also, the results for Spreading coupling suggest the existence of an optimal range of interaction strength where the network maintains its dynamical stability. The fact that the instability appearing at strong inter-node coupling remains localized to a small sub-network indicates the set of nodes/links which are crucial for the stability of this network structure. These nodes/links may be also used as a target for dynamical manipulation of the network. The unstable sub-network is dynamically linked to the stable part of the network, suggesting a complex balance of the global network's behavior, that needs additional study.

We hope that our work traces the appropriate methodology for the study of collective dynamics in coupled multi-dimensional chaotic maps with time delay on empirical networks, in particular in the context of searching for a network architecture with robust and dynamically controllable behavior.\\[0.05cm]

\n {\bf Acknowledgments.} This work was supported by the DFG through Project FOR868, by the national Program P1-0044 (Slovenia), and in part by the European Project MRTN-CT-2004-005728 (PATTERNS). Many thanks to A. D\'iaz-Guilera and M. \v Suvakov for useful discussions. Special thanks to R. Krivec for maintaining the computing resources at Dept. of Theor. Physics (J. Stefan Institute), where the numerical work was performed.


\begin{scriptsize}

\end{scriptsize}
\end{document}